\documentclass[aps,prc,reprint,amsmath,amssymb,showpacs,preprintnumbers,superscriptaddress]{revtex4-1}
\usepackage{CJK}
\usepackage{graphicx}% Include figure files
\usepackage{dcolumn}% Align table columns on decimal point
\usepackage{bm}% bold math
\usepackage{color}
\usepackage{hyperref}% add hypertext capabilities
\allowdisplaybreaks

\begin{document}
\begin{CJK*}{UTF8}{}
\title{Microscopic analysis of induced nuclear fission dynamics}
\CJKfamily{gbsn}
\author{Z. X. Ren}
\affiliation{State Key Laboratory of Nuclear Physics and Technology, School of Physics, Peking University, Beijing 100871, China}
\author{J. Zhao}
\affiliation{Center for Circuits and Systems, Peng Cheng Laboratory, Shenzhen 518055, China}
\author{D. Vretenar}
\email{vretenar@phy.hr}
\affiliation{Physics Department, Faculty of Science, University of Zagreb, 10000 Zagreb, Croatia}
\affiliation{State Key Laboratory of Nuclear Physics and Technology, School of Physics, Peking University, Beijing 100871, China}
\author{T. Nik\v si\' c}
\affiliation{Physics Department, Faculty of Science, University of Zagreb, 10000 Zagreb, Croatia}
\affiliation{State Key Laboratory of Nuclear Physics and Technology, School of Physics, Peking University, Beijing 100871, China}
\author{P. W. Zhao}
\email{pwzhao@pku.edu.cn}
\affiliation{State Key Laboratory of Nuclear Physics and Technology, School of Physics, Peking University, Beijing 100871, China}
\author{J. Meng}
\email{mengj@pku.edu.cn}
\affiliation{State Key Laboratory of Nuclear Physics and Technology, School of Physics, Peking University, Beijing 100871, China}

\begin{abstract}
The dynamics of low-energy-induced fission is explored using a consistent microscopic framework that combines the time-dependent generator coordinate method (TDGCM) and time-dependent nuclear density functional theory (TDDFT). While the former presents a fully quantum mechanical approach that describes the entire fission process as an adiabatic evolution of collective degrees of freedom, the latter  models the dissipative dynamics of the final stage of fission by propagating the nucleons independently toward scission and beyond. The two methods, based on the same nuclear energy density functional and pairing interaction, are employed in an illustrative study of the charge distribution of yields and total kinetic energy for induced fission of $^{240}$Pu. For the saddle-to-scission phase a set of initial points for the TDDFT evolution is selected along an isoenergy curve beyond the outer fission barrier on the deformation energy surface, and the TDGCM is used to calculate the probability that the collective wave function reaches these points at different times. Fission observables are computed using both methods and compared with available data.
\end{abstract}

\date{\today}

\maketitle

\end{CJK*}

\section{Introduction}

A unified microscopic framework for the description of the entire process of nuclear fission is still not available \cite{bender20,schmidt18,schunck16}. This is due to the fact that fission presents an extremely complicated quantum many-body problem but also because the time evolution of the order of 20 -- 50 zs ( 1 zs $ = 10^{-21}$ s) \cite{jacquet09} basically consists of two distinct intervals characterized by very different dynamics. The slow evolution from the quasistationary initial state to the outer fission barrier (saddle point) can be described by a relatively small number of collective degrees of freedom. Beyond the saddle point fission dynamics becomes dissipative, and the nucleus quickly elongates toward scission.

Two basic microscopic approaches to the description of induced fission dynamics have been developed. The time-dependent generator coordinate method (TDGCM) \cite{krappe12,schunck16,younes19,Regnier2016_PRC93-054611} represents the nuclear wave function by a superposition of generator states that are functions of collective coordinates, and can be applied to an adiabatic description of the entire fission process. Beyond the outer fission barrier, however, collective dynamics cannot be decoupled from intrinsic nucleon motion, and the dissipative dynamics is described by models based on time-dependent density functional theory (TDDFT)  \cite{simenel12,simenel18,nakatsukasa16,stevenson19,bulgac16,magierski17,scamps18,bulgac19,bulgac20}. However, since TDDFT-based models deal with the classical evolution of independent nucleons in  mean-field potentials, they cannot be applied in the classically forbidden region of the collective space nor do they take into account quantum fluctuations.

Therefore, on the one hand, TDGCM presents a fully quantum mechanical approach but only takes into account collective degrees of freedom in the adiabatic approximation. On the other hand, nuclear TDDFT automatically includes the one-body dissipation mechanism, but can only simulate a single fission event by propagating the nucleons independently. The relative importance of these effects when calculating fission observables, such as fission yields or the kinetic energy distribution, has been discussed in many studies but never compared quantitatively in a consistent way. In fact, already more than 40 years ago \cite{negele78} it was suggested that a description of the entire fission process could be realized by using an adiabatic model for the time interval in which the fissioning nucleus evolves from the quasistationary initial state to the saddle point and nonadiabatic method for the saddle-to-scission and beyond-scission dynamics (see also Ref.~\cite{simenel14}). We note that, along these lines, a semiphenomenological approach was adopted in Ref.~\cite{sadhukhan16} to compute the distribution of spontaneous fission yields of $^{240}$Pu.
In the classically forbidden region WKB was used to calculate a family of fission probabilities that correspond to the hypersurface of the outer turning points. The potential energy surface and collective inertia were obtained using nuclear DFT. Starting from the outer turning points, the time dependent fission paths to scission were then computed by solving the dissipative Langevin equations.

In the present study we combine the TDGCM and TDDFT in a consistent microscopic framework to analyze the final stage of the fission process in which, after the nucleus has passed over the saddle point, it deforms toward scission. More precisely, a set of initial points for the TDDFT evolution is selected along an isoenergy curve beyond the outer fission barrier on the deformation energy surface, and the TDGCM is used to calculate the probability that the collective wave function reaches these points at different times. Both the TDGCM and TDDFT are then used to calculate the fission yields and kinetic energy distribution.
The particular implementations of the TDGCM and TDDFT used in this work can be found in Refs. \cite{Tao2017_PRC96-024319, Zhao2019_PRC99-014618, Zhao2019_PRC99-054613} and \cite{ren20LCS, ren20O16}, respectively.
Both models are based on the relativistic energy density functional PC-PK1 \cite{zhao10} and a monopole pairing interaction with the Bardeen-Cooper-Schrieffer (BCS) approximation~\cite{Ring2004manybody, Bender2000pairing}.

\section{Theoretical framework}\label{sec_theo}
\subsection{The time-dependent generator coordinate method plus Gaussian overlap approximation}

The time-dependent generator coordinate method plus Gaussian overlap approximation (TDGCM+GOA) describes induced fission as a slow adiabatic process determined by a small number of  collective degrees of freedom. Nonadiabatic effects arising from the coupling between collective and intrinsic degrees of freedom are not taken into account. Fission dynamics is governed by a local, time-dependent Schr\"odinger-like equation
in the space of collective coordinates $\bm{q}$:
\begin{equation}
i\hbar \frac{\partial g(\bm{q},t)}{\partial t} = \hat{H}_{\rm coll} (\bm{q}) g(\bm{q},t) ,
\label{eq:TDGCM}
\end{equation}
where $g(\bm{q},t)$ is the complex wave function of the collective variables $\bm{q}$ and time $t$.
Axial symmetry is assumed with respect to the axis along which the two fission fragments eventually separate, and a two-dimensional (2D) collective space of quadrupole $\beta_{20}$ and octupole $\beta_{30}$ deformation parameters is considered.
The collective Hamiltonian $\hat{H}_{\rm coll} (\bm{q})$ reads
\begin{align}
\hat{H}_{\rm coll} (\beta_{20},\beta_{30}) = &- {\hbar^2 \over 2}
\sum_{ij =2,3} {\partial \over \partial \beta_i} B_{ij}(\beta_{20},\beta_{30}) {\partial \over \partial \beta_j}\nonumber \\
&+ V(\beta_{20},\beta_{30}),
\label{eq:Hcoll2}
\end{align}
where $B_{ij}(\beta_{20},\beta_{30})$ and $V(\beta_{20},\beta_{30})$ denote the inertia tensor and collective potential, respectively.
The inertia tensor is the inverse of the mass tensor, $B_{ij}(\beta_{20},\beta_{30}) =(\mathcal{M}^{-1})_{ij}$.
The mass tensor is calculated in the perturbative cranking approximation~\cite{Baran11GCM}
\begin{equation}
\label{eq:pmass}
\mathcal{M}^{Cp} = \hbar^2 {\it M}_{(1)}^{-1} {\it M}_{(3)} {\it M}_{(1)}^{-1},
\end{equation}
where
\begin{equation}
\label{eq:mmatrix}
\left[ {\it M}_{(k)} \right]_{ij} = \sum_{\mu\nu}
    {\langle 0 | \hat{Q}_i | \mu\nu \rangle
     \langle \mu\nu | \hat{Q}_j | 0 \rangle
     \over (E_\mu + E_\nu)^k}.
\end{equation}
$|\mu\nu\rangle$ are two-quasiparticle states and $E_\mu$, $E_\nu$ denote the corresponding quasiparticle energies.

The input for the calculation of the collective mass, that is, the single-quasiparticle states, energies, and occupation factors, are calculated in
a self-consistent mean-field approach based on nuclear energy density functionals.
The map of the energy surface as function of the quadrupole and octupole deformations is obtained by imposing constraints on the corresponding mass moments:
\begin{equation}
\label{eq:multipole-moments}
\hat{Q}_{20} = 2z^2 - r_\perp^2 \quad \textnormal{and} \quad \hat{Q}_{30} = 2z^3 - 3z r_\perp^2.
\end{equation}
The deformation parameters $\beta_{20}$ and $\beta_{30}$ are determined using the following relations:
\begin{equation}
\beta_{20} = \frac{\sqrt{5\pi}}{3AR_0^2} \langle \hat{Q}_{20} \rangle \quad \textnormal{and} \quad
\beta_{30} = \frac{\sqrt{7\pi}}{3AR_0^3} \langle \hat{Q}_{30} \rangle,
\end{equation}
with $R_0=r_0A^{1/3}$ and $r_0=1.2$ fm.
The collective potential $V(\beta_{20},\beta_{30})$ is obtained by subtracting the vibrational zero-point energy (ZPE) from the total
mean-field energy~\cite{Girod79GCM}
\begin{equation}
\label{eq:zpe}
E_{\rm ZPE} = {1\over4} {\rm Tr} \left[ {\it M}_{(3)}^{-1} {\it M}_{(2)} \right],
\end{equation}
where the ${\it M}_{(k)}$ are given by Eq.~(\ref{eq:mmatrix}).

The collective space is divided into an inner region with a single nuclear density distribution,
and an external region that contains two separated fission fragments.
The set of configurations that divides the inner and external regions defines the scission
hypersurface.  The flux of the probability current through this
hypersurface provides a measure of the probability of observing a given pair of fragments at time $t$.
Each infinitesimal surface element is associated with a given pair of fragments $(Z_L, Z_H)$, where $Z_L$ and $Z_H$ denote the
charge of the lighter and heavier fragments, respectively.
The integrated flux $F(\xi,t)$ for a given surface element $\xi$ is defined as \cite{regnier18}
\begin{equation}
F(\xi,t) = \int_{t_0}^{t} dt^\prime \int_{\{ \beta_{20}, \beta_{30} \}\in \xi} \bm{J}(\beta_{20},\beta_{30},t^\prime) \cdot d\bm{S},
\label{eq:flux}
\end{equation}
where $\bm{J}(\beta_{20},\beta_{30},t)$ is the current
\begin{align}
\label{eq:current}
J_k(\beta_{20},\beta_{30},t) =\,&{\hbar \over 2i} \bm{B}^{-1}(\bm{q}) [g^{*}(\bm{q},t) \nabla g(\bm{q},t) \\
&- g(\bm{q},t) \nabla g^{*}(\bm{q},t)].
\end{align}
The yield for the fission fragment with charge $Z$ is defined by
\begin{equation}
Y(Z) \propto \sum_{\xi \in \mathcal{A}} \lim_{t \rightarrow \infty} F(\xi,t).
\end{equation}
The set $\mathcal{A}(\xi)$ contains all elements belonging to the scission hypersurface such that one of the fragments has charge number $Z$.

In the present study the mean-field deformation energy is calculated with the multidimensionally constrained relativistic mean-field (MDC-RMF) model \cite{Lu12MDCRMF,Lu14MDCRMF,zhou16MDCRMF,zhao16MDCRMF}, and
calculations are performed using the point-coupling relativistic energy density functional PC-PK1~\cite{zhao10}.
Pairing correlations are taken into account in the BCS approximation~\cite{Ring2004manybody} with a monopole pairing interaction.
The cutoff function for the pairing window is the same as in Ref.~\cite{Scamps13TDBCS}.
The pairing strength parameters: $-0.135$ MeV for neutrons, and $-0.230$ MeV for protons, are determined by the empirical pairing gaps of $^{240}$Pu,
using the three-point odd-even mass formula~\cite{Bender2000pairing}.
The mean-field equations are solved by expanding the nucleon Dirac spinors in the axially deformed harmonic oscillator (ADHO)
basis with $N_f=20$ oscillator shells. Reference~\cite{Lu14MDCRMF} details the multidimensionally constrained relativistic mean-field model.

The fission process is described by the time evolution of an initial wave packet $g(\bm{q},t=0)$ ($\bm{q} \equiv \{\beta_{20},\beta_{30}\}$),
built as a Gaussian superposition of the quasibound states $g_k$,
\begin{equation}
g(\bm{q},t=0) = \sum_{k} \exp\left( { (E_k - \bar{E} )^{2} \over 2\sigma^{2} } \right) g_{k}(\bm{q}),
\label{eq:initial-state}
\end{equation}
where the value of the parameter $\sigma$ is set to 0.5 MeV. The collective states $\{ g_{k}(\bm{q}) \}$
are solutions of the stationary eigenvalue equation in which the original collective potential $V(\bm{q})$ is replaced by a
new potential $V^{\prime} (\bm{q})$ that is obtained by extrapolating the inner potential barrier with a quadratic form.
The mean energy $\bar{E}$ in Eq.~(\ref{eq:initial-state}) is then adjusted iteratively in
such a way that $\langle g(t=0)| \hat{H}_{\rm coll} | g(t=0) \rangle = E_{\rm coll}^{*}$, and this
average energy $E_{\rm coll}^{*}$ is chosen $\approx 1$ MeV above the fission barrier.
The TDGCM+GOA Hamiltonian of Eq.~(\ref{eq:Hcoll2}), with the original collective potential
$V(\bm{q})$, propagates the initial wave packet in time.
The computer code employed for modeling the time evolution of the fissioning nucleus is FELIX (version 2.0) \cite{regnier18}.
The time step is $\delta t=5\times 10^{-4}$ zs (1 zs $= 10^{-21}$ s), and the charge
distributions are calculated after $6 \times 10^{4}$ time steps, which correspond to 30 zs.
As in our recent calculations of Refs.~\cite{Tao2017_PRC96-024319,Zhao2019_PRC99-014618,Zhao2019_PRC99-054613,zhao20,zhao21},
the parameters of the additional imaginary absorption potential that takes into account the escape
of the collective wave packet  in the domain outside the region of calculation \cite{regnier18} are:
the absorption rate $r=20\times 10^{22}$ s$^{-1}$ and the width of the absorption band $w=1.0$.
The scission contour that divides the inner and external regions is determined by the Gaussian neck operator $\displaystyle \hat{Q}_{N}=\exp[-(z-z_{N})^{2} / a_{N}^{2}]$,
where $a_{N}=1$ fm and $z_{N}$ is the position of the neck~\cite{Younes2009fission}.
In this work we define the prescission domain by $\langle \hat{Q}_{N} \rangle>4$, and consider the frontier of this domain as the scission contour.

\subsection{Time-dependent covariant density functional theory}
The dissipative dynamics of the saddle-to-scission phase of the fission process is modeled with the time-dependent covariant DFT~\cite{ren20LCS, ren20O16}.
Pairing correlations are treated dynamically with the time-dependent BCS approximation~\cite{ebata10TDBCS, Scamps13TDBCS}.
The wave function of the system takes the general form of a quasiparticle vacuum,
\begin{equation}
   |\Psi(t)\rangle = \prod_{k>0}\left[u_k(t)+v_k(t)c_k^+(t)c_{\bar{k}}^+(t)\right]|0\rangle,
\end{equation}
where $u_k(t)$ and $v_k(t)$ are the parameters in the transformation between the canonical and the quasiparticle states,
and $c_k^+(t)$ stands for the creation operator associated with the canonical state $\psi_k(\bm{r},t)$.
The evolution of $\psi_k(\bm{r},t)$ is determined by the time-dependent Dirac equation
\begin{equation}\label{Eq_td_Dirac_eq_BCS}
  i\frac{\partial}{\partial t}\psi_k(\bm{r},t)=\left[\hat{h}(\bm{r},t)-\varepsilon_k(t)\right]\psi_k(\bm{r},t),
\end{equation}
where the single-particle energy $\varepsilon_k(t)=\langle\psi_k|\hat{h}|\psi_k\rangle$, and the single-particle Hamiltonian $\hat{h}(\bm{r},t)$ reads
\begin{equation}
   \hat{h}(\bm{r},t) = \bm{\alpha}\cdot(\hat{\bm{p}}-\bm{V})+V^0+\beta(m_N+S).
\end{equation}
The scalar $S(\bm{r},t)$ and four-vector $V(\bm{r},t)$ potentials are consistently determined at each step in time by the time-dependent densities and currents in the isoscalar-scalar, isoscalar-vector and isovector-vector channels,
\begin{subequations}\label{Eq_density_current}
  \begin{align}
    &\rho_s(\bm{r},t)=\sum_kn_k\bar{\psi}_k\psi_k,\\
    &j^\mu(\bm{r},t)=\sum_kn_k\bar{\psi}_k\gamma^\mu\psi_k,\\
    &j_{TV}^\mu(\bm{r},t)=\sum_kn_k\bar{\psi}_k\gamma^\mu\tau_3\psi_k,
  \end{align}
\end{subequations}
respectively. $\tau_3$ is the isospin Pauli matrix (for details, see Ref.~\cite{ren20O16}). The time evolution of the
occupation probability $n_k(t)=|v_k(t)|^2$, and pairing tensor $\kappa_k(t)=u_k^*(t)v_k(t)$, is governed by the following equations
\begin{subequations}\label{Eq_td_nkapp_eq_BCS}
   \begin{align}
     &i\frac{d}{dt}n_k(t)=n_k(t)\Delta_k^*(t)-n_k^*(t)\Delta_k(t),\\
     &i\frac{d}{dt}\kappa_k(t)=[\varepsilon_k(t)+\varepsilon_{\bar{k}}(t)]\kappa_k(t)+\Delta_k(t)[2n_k(t)-1].
   \end{align}
\end{subequations}
In time-dependent calculations, a monopole pairing interaction is employed, and the gap parameter $\Delta_k(t)$ is determined by the single-particle energy and pairing tensor,
\begin{equation}
  \Delta_k(t)=\left[G\sum_{k'>0}f(\varepsilon_{k'})\kappa_{k'}\right]f(\varepsilon_k),
\end{equation}
where $f(\varepsilon_k)$ is the cutoff function for the pairing window.

In calculations with time-dependent covariant DFT, the mesh spacing of the lattice is 1.0 fm for all directions, and the box size is taken as $L_x\times L_y\times L_z=20\times20\times60~{\rm fm}^3$.
The time-dependent Dirac Eq. \eqref{Eq_td_Dirac_eq_BCS} is solved with the predictor-corrector method, and the time-dependent Eqs. \eqref{Eq_td_nkapp_eq_BCS} using the Euler algorithm.
The step for the time evolution is $6.67\times10^{-4}$~zs.
The density functional, pairing strength parameters $G$, and the cutoff function $f(\varepsilon_k)$ for the pairing window are the same as in the corresponding TDGCM calculation.
The initial states for the time evolution are obtained by self-consistent deformation-constrained relativistic DFT calculations in a three-dimensional lattice space, using the inverse Hamiltonian and Fourier spectral methods~\cite{ren17dirac3d, ren19LCS, ren20_NPA}, with the box size: $L_x\times L_y\times L_z=20\times20\times50~{\rm fm}^3$.

\section{Induced fission dynamics of $^{240}$Pu}
%-------------------------------------------------------------------------------------
\begin{figure}[tbh!]
\centering
\includegraphics[width=0.45\textwidth]{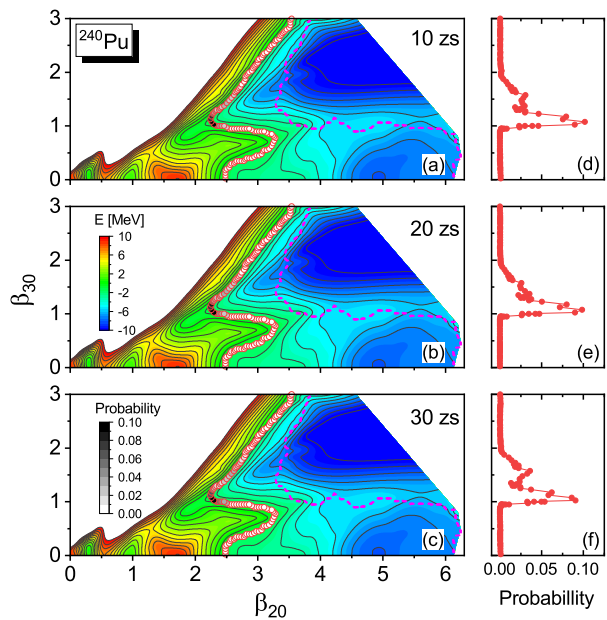}\\
\caption{Self-consistent deformation energy surface of $^{240}$Pu in the plane of quadrupole-octupole axially symmetric deformation parameters, calculated with the relativistic density functional PC-PK1 and a monopole pairing interaction. Contours join points on the surface with the same energy, and the contour interval is 1 MeV. The open dots correspond to points on the isoenergy curve 1 MeV below the energy of the equilibrium minimum. The color code of the dots and the corresponding panel on the right, denote the normalized probability that the initial TDGCM wave packet reaches the particular point after 10 zs (top), 20 zs (middle), and 30 zs (bottom). The dashed curve denotes the scission contour, defined by the expectation value of the Gaussian neck operator $\langle \hat{Q}_{N} \rangle=4$.}
 \label{fig:240PES}
\end{figure}
%-------------------------------------------------------------------------------------
In Fig.~\ref{fig:240PES} we display the two-dimensional microscopic self-consistent mean-field deformation energy surface of $^{240}$Pu, as a function of the axial quadrupole ($\beta_{20}$) and octupole ($\beta_{30}$) deformation parameters,
calculated with the multidimensionally constrained relativistic mean-field (MDC-RMF) model~\cite{Lu12MDCRMF, Lu14MDCRMF, zhou16MDCRMF, zhao16MDCRMF}.
The range for the collective variable $\beta_{20}$ is $0 \le \beta_{20} \le 7$ with
a step $\Delta \beta_{20} = 0.04$, while the collective variable $\beta_{30}$ is considered in the interval $0 \le \beta_{30} \le 3.5$ with a step $\Delta \beta_{30} =0.05$.
The equilibrium minimum is calculated at $\beta_{20}\approx 0.3$ and $\beta_{30}=0$, and the isomeric minimum at $\beta_{20}\approx 0.9$ and $\beta_{30}=0$, in agreement with empirical values. One also notes the two fission barriers, and the fission valley at large deformations.

The open dots on the energy surface in Fig.~\ref{fig:240PES} correspond to the isoenergy curve 1 MeV below the energy of the equilibrium minimum, located beyond the outer fission barrier.
These points will be used as initial locations for the TDDFT calculation.
The color code of these dots, as well as the corresponding panel on the right, denote the probability that the initial TDGCM wave packet reaches the particular point after a specific time. This is, of course, just the square modulus of the collective wave function, and we display these probabilities after 10 zs (top), 20 zs (middle), and 30 zs (bottom). The probability (normalized to 1 at each time) appears concentrated in the region $\beta_{20}\approx 2.2$ and $\beta_{30} \approx 1$ -- $1.5$.

\begin{figure}[tbh!]
\centering
\includegraphics[width=0.45\textwidth]{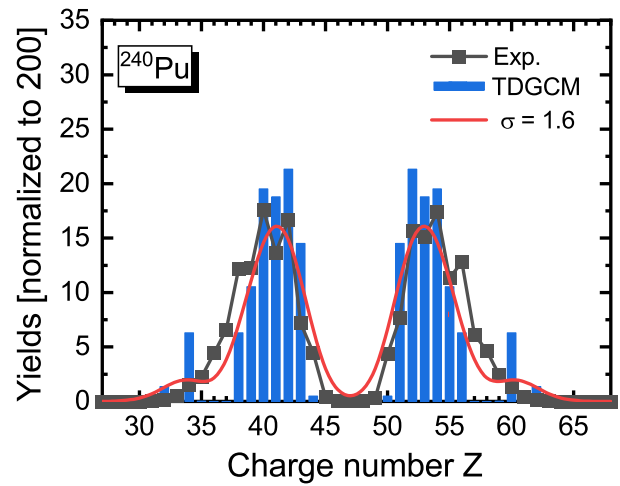}\\
 \caption{Charge yields for induced fission of $^{240}$Pu. The calculated fission yields (solid red curve) are obtained by convoluting the raw flux (blue bars) with a Gaussian function of the number of particles, using a width of 1.6 units. The data are from Ref.~\cite{ramos18} and correspond to an average excitation energy of 10.7 MeV.}
 \label{fig:yields}
\end{figure}

In Fig.~\ref{fig:yields} we plot the charge yields obtained with the TDGCM, normalized to $\sum_{Z} Y(Z) = 200$, in comparison to the experimental fragment charge distribution \cite{ramos18}. The calculated fission yields are obtained by convoluting the raw flux with a Gaussian function of the number of particles, and the width is 1.6 units.
The TDGCM calculation reproduces the trend of the data except, of course, the odd-even staggering. The predicted asymmetric peaks are located at $Z = 41$ and $Z = 53$, one mass unit away from the experimental peaks, and we note that the model does not quantitatively reproduce the asymmetric tails of the empirical distribution. However, considering that no additional adjustment has been made for the parameters of the model, the agreement with experiment is very good.

%-------------------------------------------------------------------------------------
%-------------------------------------------------------------------------------------

To calculate charge yields with the TDDFT approach, we employ the time-dependent relativistic (covariant) DFT in a three-dimensional lattice space, with pairing correlations  treated dynamically in the time-dependent BCS approximation~\cite{ebata10TDBCS, Scamps13TDBCS}.
The evolution of single-particle wave functions, and pairing factors $n_k(t)$ and $\kappa_k(t)$, is governed by Eqs.~\eqref{Eq_td_Dirac_eq_BCS} and \eqref{Eq_td_nkapp_eq_BCS}.

\begin{figure}[tbh!]
\centering
 \includegraphics[width=0.45\textwidth]{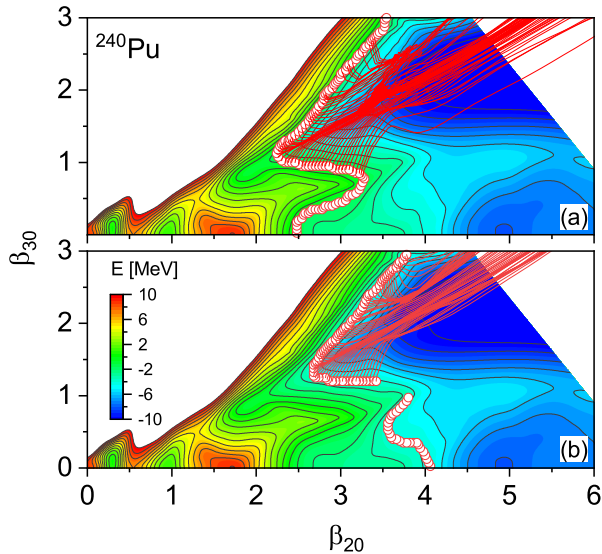}
 \caption{TD(C)DFT fission trajectories from the initial points (denoted by open dots ) on the selfconsistent deformation energy surface of $^{240}$Pu.
 The initial points in the upper (lower) panel correspond to the iso-energy contours at $-1$ MeV ($-4$ MeV ) below the energy of the equilibrium minimum.
 Only those trajectories that end up in scission of the fissioning nucleus are shown.
Trajectories that start from very asymmetric shapes (large $\beta_{30}$ values in the upper panel), or from almost symmetric shapes (small $\beta_{30}$ values in both panels), do not lead to scission but get trapped in local minima. The total number of trajectories included in the figure is 133 (62 for the initial energy $E = -1$ MeV, and 71 for $E = -4$ MeV).
 }
 \label{fig3}
\end{figure}

Given the initial single-nucleon wave functions and occupation probabilities, determined in a mean-field approach with constraints on the collective coordinates in three-dimensional lattice space~\cite{ren17dirac3d, ren19LCS, ren20_NPA}, TD(C)DFT propagates the nucleons independently toward scission. This method cannot be used to model the slow evolution from the equilibrium deformation to the saddle point and, therefore, the starting point is usually taken below the outer barrier. However, if this point is too close to the barrier, the trajectory can get confined in a region of a local minimum. The set of initial points that we choose in the first example (open dots in Fig.~\ref{fig:240PES}), corresponds to a deformation energy of 1 MeV below the energy of the equilibrium minimum.
In Fig.~\ref{fig3}(a) we plot the TD(C)DFT fission trajectories from the initial points (denoted by open dots) on the self-consistent deformation energy surface of $^{240}$Pu.
It can be seen that most trajectories simply follow the path of steepest descent.

%-------------------------------------------------------------------------------------
%-------------------------------------------------------------------------------------

%-------------------------------------------------------------------------------------
\begin{figure}[tbh!]
\centering
\includegraphics[width=0.45\textwidth]{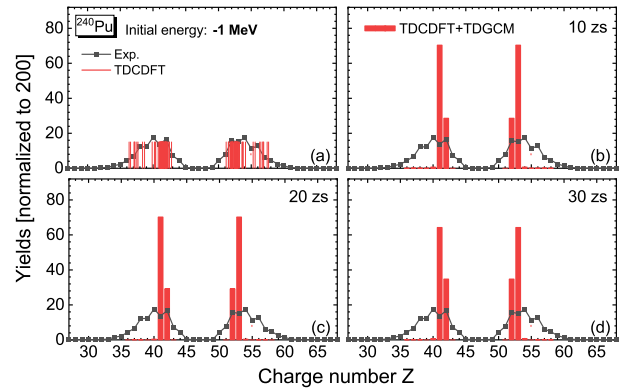}\\
 \caption{(a) The $^{240}$Pu fission charge fragments calculated with the TD(C)DFT, starting from the initial points on the isoenergy curve beyond the outer barrier, 1 MeV bellow the equilibrium minimum (cf. Fig.~\ref{fig:240PES}). (b - d) The corresponding charge yields, for the three cases when the TD(C)DFT calculation is initiated 10, 20, and 30 zs, respectively, after the initial TDGCM wave packet starts propagating from the equlibrium minimum. The data are from Ref.~\cite{ramos18} and correspond to an average excitation energy of 10.7 MeV.}
\label{fig:TDCDFT_1}
\end{figure}
%-------------------------------------------------------------------------------------

In the upper left panel of Fig.~\ref{fig:TDCDFT_1}, we plot the resulting TD(C)DFT charge fragments, as well as the experimental charge yields. The vertical bars do not represent the charge yields but rather denote the light and heavy fragments that are obtained for a particular trajectory.
While the TDGCM collective wave function sweeps the entire energy surface and the flux through any element of the scission hypersurface can be calculated, in TD(C)DFT a single fission event is obtained following a trajectory that starts from a given initial point.
However, not all the TD(C)DFT trajectories that start from the points shown in Fig.~\ref{fig:240PES} lead to scission [cf. Fig.~\ref{fig3}(a)].
One notices that most scission events are obtained in the intervals $Z = 40$ -- $42 $ and $Z = 52$ -- $54$, in agreement with experiment \cite{ramos18}. To calculate the charge yields, we multiply each scission event by the probability that the collective wave packet has reached  the corresponding point after a specific time (panels on the right of Fig.~\ref{fig:240PES}) and, as in the TDGCM calculation, normalize the yields to 200. When compared to the data, it appears that the TD(C)DFT yields qualitatively reproduce the position of the peaks but not the tails of the experimental distribution. As shown in the panels (b), (c), and (d), this result basically does not depend on the instant (10, 20, or 30 zs) when the TD(C)DFT calculation was initiated.

\begin{figure}[tbh!]
\centering
\includegraphics[width=0.45\textwidth]{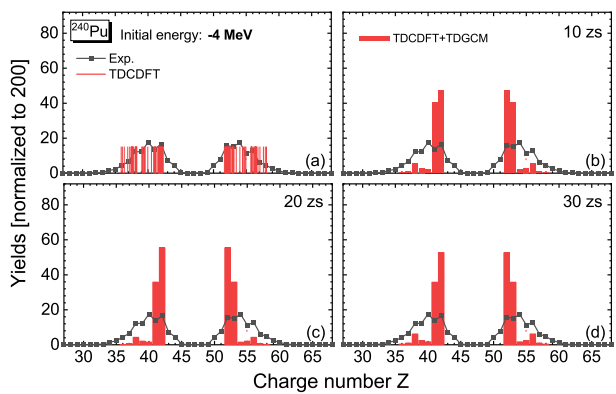}\\
 \caption{Same as in the caption to Fig.~\ref{fig:TDCDFT_1} but for the initial isoenergy curve 4 MeV bellow the energy of the equilibrium minimum.}
 \label{fig:TDCDFT_2}
\end{figure}

\begin{figure}[tbh!]
\centering
\includegraphics[width=0.45\textwidth]{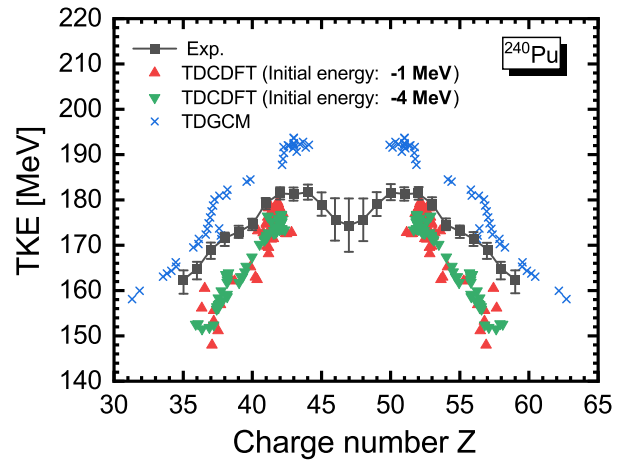}\\
 \caption{The calculated total kinetic energies of the nascent fragments for induced fission of $^{240}$Pu, as functions of the fragment charge. The TDGCM and TD(C)DFT results are shown in comparison to the data \cite{Caamano2015}.
}
 \label{fig:TKE}
\end{figure}

Similarly, Fig.~\ref{fig:TDCDFT_2} displays the results of the same TD(C)DFT calculation, but now starting from a set of initial points on the isoenergy curve that is 4 MeV below the energy of the equilibrium minimum.
The TD(C)DFT fission trajectories are shown in Fig.~\ref{fig3}(b), where the disconnected region without open dots in the lower panel correspond to points on the deformation energy surface that, in the TDGCM calculation, are located already beyond the scission contour defined by the number of particles in the neck.
In this case the TD(C)DFT evolution starts closer to scission, and we note that more fission events are obtained in the tails of the distribution.
The corresponding yields [panels (b) to (d)] also exhibit a somewhat richer structure and, as in the previous case, essentially do not depend on the initial time of the TD(C)DFT evolution.
An interesting result is obtained when the TD(C)DFT charge yields shown in Figs.~\ref{fig:TDCDFT_1} and \ref{fig:TDCDFT_2}, are compared with those obtained using the TDGCM (cf. Fig.~\ref{fig:yields}), and with the data. Obviously, the TDGCM does a better job in reproducing the empirical charge yields.  This means that the fragment distribution is already determined before the final stage of the fission process in which the dissipation mechanism becomes important \cite{zhang16}. The TD(C)DFT reproduces the peaks of the experimental charge yields but not the width. Only when the set of initial points on the deformation energy surface is located much closer to the fission valley, the calculated fission yields exhibit a structure that qualitatively resembles the empirical charge yields. This emphasizes the importance of quantum fluctuations that are included in the TDGCM evolution of the collective nuclear function, but not in the TD(C)DFT trajectories that correspond to the propagation of individual nucleons in mean-field potentials.
A different result is obtained for the total kinetic energy (TKE) of the fragments. In Fig.~\ref{fig:TKE} we show the TKEs of the nascent fission fragments for
$^{240}$Pu, as functions of the fragment charge. The theoretical values are compared to data \cite{Caamano2015}. In the TDGCM, the total kinetic energy for a particular pair of fragments can be evaluated from
\begin{equation}
E_{\rm TKE}=\frac{e^2Z_H Z_L}{d_{\rm ch}},
\label{eq:TKE}
\end{equation}
where $e$ is the proton charge, $Z_H (Z_L)$ the charge of the heavy (light) fragment, and $d_{\rm ch}$  is the distance between centers of charge at the point of scission.
For TD(C)DFT, the TKE at a finite distance between the fission fragments ($\approx 25$ fm, at which shape relaxation brings the fragments to their equilibrium shapes) is calculated using the expression \cite{bulgac19}
\begin{equation}
E_\textrm{TKE} = \frac{1}{2}mA_{\mathrm{H}} \bm{v}_{\mathrm{H}}^2 + \frac{1}{2}mA_{\mathrm{L}} \bm{v}_{\mathrm{L}} ^2 + E_{\mathrm{Coul}},
\label{eq:TKE-TDDFT}
\end{equation}
where the velocity of the fragment $f = H, L$ reads
\begin{equation}
\vec{v}_f = \frac{1}{mA_f} \int_{V_f} d{\bf r}~\bm{j}(\bm{r}),
\end{equation}
and $\bm{j}(\bm{r})$ is the total current density. The integration is over the half-volume corresponding to the fragment $f$, and $E_{\mathrm{Coul}}$ is the Coulomb energy.

%-------------------------------------------------------------------------------------
%-------------------------------------------------------------------------------------

%-------------------------------------------------------------------------------------
%-------------------------------------------------------------------------------------

TDGCM by definition describes nondissipative dynamics and, in the adiabatic approximation, all the potential energy is converted into collective kinetic energy during the saddle-to-scission evolution. The nascent fragments are cold, and the calculated TKEs are systematically too large. However, one-body dissipation is automatically included in TD(C)DFT and, in the short time interval it takes from the initial point to scission, the collective flow energy is converted into intrinsic degrees of freedom and the nucleus heats up \cite{bulgac19}. This results in a lower TKE, as show in Fig.~\ref{fig:TKE}. In addition, because of shape relaxation after scission, the deformation energy of the fragments is also converted into internal heat. It is interesting to note that the calculated TKEs essentially do not depend on whether we chose the initial points at 1 or 4 MeV below the energy of the equilibrium minimum.

It appears that TD(C)DFT slightly underestimates the TKE for the fragments close to the peaks of the charge yields distribution but predicts TKEs considerably below the experimental values for the tails of the distribution. Similar results for the TKE of $^{240}$Pu fragments were also obtained in the TDDFT study of Ref.~\cite{scamps18}. We note that the values calculated using Eq.~(\ref{eq:TKE-TDDFT}) present a lower bound for the total kinetic energy, due to the fact that this expression does not include the contribution of prescission energy. Namely, while for the TDGCM the average energy of the initial wave packet $E_{\rm coll}^{*}$ is chosen $1$ MeV above the fission barrier ($\approx 8$ MeV for $^{240}$Pu), all the initial points for the TDDFT calculation are on the deformation energy surface, 1 or 4 MeV below the energy of the equilibrium minimum. Thus, the starting points for TDDFT trajectories are more than 10 MeV below the `physical' value. However, one cannot simply add this difference to the TKE, because part of the prescission energy will be converted into excitation energy of the nascent fragments. It is not possible to give a quantitative estimate for the portion of the prescission energy that will be converted into TKE but, in any case, this contribution will increase the values shown in Fig.~\ref{fig:TKE}.

\section{Summary}\label{sec_summ}
In summary, a consistent microscopic framework, based on TDGCM and TDDFT, has been applied to model the entire process of induced fission. Given a nuclear energy density functional and pairing interaction, the TDGCM is used to evolve adiabatically a set of collective degrees of freedom of the fissioning system from the quasistationary initial state to the outer barrier and beyond. Starting from a isoenergy contour below the outer barrier, for which the TDGCM provides the probabilities that the collective wave functions reaches these points at any given time, the TDDFT is used to model the dissipative fission dynamics in the saddle-to-scission phase. By combining the two methods, an illustrative study has been performed of the charge distribution of yields for low-energy induced fission of $^{240}$Pu.

Even though this type of approach to fission dynamics was suggested more than forty years ago, it has only been quantitatively tested for the first time in the present study. The results obtained for $^{240}$Pu, indicate that the TDGCM+TDDFT method is, in fact, less than optimal. Quantum fluctuations, included in TDGCM but not in TDDFT, are essential for a quantitative estimate of fission yields. Dissipative effects, taken into account in TDDFT but not in TDGCM, are crucial for the total kinetic energy distribution. Even when the two methods are combined, the weak points of each approach cannot be removed completely.

This work indicates a direction in which microscopic modeling of fission based on TDDFT can be expanded further. A number of open questions will be investigated in forthcoming studies, that will involve a larger set of nuclei, different functionals, different methods for the calculation of collective inertia \cite{zhao20}, and dynamical treatment of pairing correlations \cite{zhao21}. The TDGCM+TDDFT approach will complement recent efforts to develop a unified microscopic framework for fission dynamics \cite{qiang21,verriere21,marevic21}.

\begin{acknowledgments}
This work has been supported in part by the High-end Foreign Experts Plan of China,
National Key R\&D Program of China (Contracts No. 2018YFA0404400 and No. 2017YFE0116700),
the National Natural Science Foundation of China (Grants No. 12070131001, No. 11875075, No. 11935003, No. 11975031, and No. 12141501),
the China Postdoctoral Science Foundation under Grant No. 2020M670013,
the High-performance Computing Platform of Peking University,
the QuantiXLie Centre of Excellence, a project co-financed by the Croatian Government and European Union through the European Regional Development Fund -- the Competitiveness and Cohesion Operational Programme (No. KK.01.1.1.01.0004),
 and the Croatian Science Foundation under the project Uncertainty quantification within the nuclear energy density framework (No. IP-2018-01-5987).
J. Z. acknowledges support by the National Natural Science Foundation of China under Grants No. 12005107 and No. 11790325.
\end{acknowledgments}

\end{document}